\documentclass{article}

\usepackage{PRIMEarxiv}

\usepackage[utf8]{inputenc} 
\usepackage[T1]{fontenc}    
\usepackage{hyperref}       
\usepackage{url}            
\usepackage{booktabs}       
\usepackage{amsfonts}       
\usepackage{nicefrac}       
\usepackage{microtype}      
\usepackage{lipsum}
\usepackage{fancyhdr}       
\usepackage{graphicx}       
\graphicspath{{media/}}     

\usepackage{multirow}
\title{Encoding feature supervised UNet++: Redesigning Supervision for liver and tumor segmentation

}

\author{
  Jiahao Cui (co-first author) \\
  Department of Mathematics \\
  College of Sciences \\
  Shanghai University\\
  Shanghai, China\\
  \texttt{cuijiahao@shu.edu.cn} \\
   \And
 Ruoxin Xiao* (co-first author) \\
  Department of Mathematics \\
  College of Sciences \\
  Shanghai University\\
  Shanghai, China\\
  \texttt{xiaoruoxin@shu.edu.cn} \\
   \And
  Shiyuan Fang\\
  School of Cyber Science and Engineering \\
  Shanghai Jiao Tong University\\
  Shanghai, China\\
  \texttt{FangShiyuanFir@sjtu.edu.cn} \\
   \And
  Minnan Pei\\
  School of Electronic \& Electrical \\
  and Communication Engineering\\
  University of Chinese Academy of Sciences\\
  Peking, China \\
  \texttt{peiminnan19@mails.ucas.ac.cn} \\
  \And
  Yixuan Yu\\
  Department of Science\\
  Northeastern University\\
  Shenyang, China \\
  \texttt{yuyixuan834@163.com} \\}
  
\begin{document}
\maketitle

\begin{abstract}
Liver tumor segmentation in CT images is a critical step in the diagnosis, surgical planning and postoperative evaluation of liver disease. An automatic liver and tumor segmentation method can greatly relieve physicians of the heavy workload of examining CT images and better improve the accuracy of diagnosis. In the last few decades, many modifications based on U-Net model have been proposed in the literature. However, there are relatively few improvements for the advanced UNet++ model.  In our paper, we propose an encoding feature supervised UNet++(ES-UNet++) and apply it to the liver and tumor segmentation. ES-UNet++ consists of an encoding UNet++ and a segmentation UNet++. The well-trained encoding UNet++ can extract the encoding features of label map which are used to additionally supervise the segmentation UNet++. By adding supervision to the each encoder of segmentation UNet++, U-Nets of different depths that constitute UNet++ outperform the original version by average 5.7\% in dice score and the overall dice score is thus improved by 2.1\%. ES-UNet++ is evaluated with dataset LiTS, achieving 95.6\% for liver segmentation and 67.4\% for tumor segmentation in dice score. In this paper, we also concluded some valuable properties of ES-UNet++ by conducting comparative anaylsis between ES-UNet++ and UNet++:(1) encoding feature supervision can accelerate the convergence of the model.(2) encoding feature supervision enhance the effect of model pruning by achieving huge speedup while providing pruned models with fairly good performance.
\end{abstract}

\keywords{UNet++ \and Encoding \and Deep Supervision \and Liver tumor \and Segmentation}

\section{Introduction}
A fully convolutional network and U-Net are the most common and essential networks in semantic segmentation, especially medical image segmentation. A fully convolutional network classifies images at the pixel level, which can accept and handle input images of any size rather than obtaining fixed-length feature vectors by fully connected layers used in classical convolutional neural networks. U-Net always has an encoder-decoder structure, with the left side for feature extraction and the right side for upsampling. U-Net can use the valid annotated data efficiently via data enhancement from a small number of training images. In addition, U-Net adopts concatenation as the way of feature confusion at the channel level. 

U-Net has various advantages. Initially, the five pooling layers enable the network to recognize the multiscale features of image features. Additionally, the upsampling part will fuse the output of the feature extraction part. However, it also has two limitations. First of all, the network runs very slowly. For each area, the network has to run once, and for the overlapping part, the network will repeat the operation. Furthermore, the network must strike a balance between accurate positioning and the acquisition of context information. The larger the patch, the more maximum pooling layers are required, which will reduce the accuracy of positioning, while the small field makes the network obtain less context information.

In our paper, we first have an introduction of our work and related work. Then our main work is proposed: The encoding feature supervised U-net++(ES-UNet++), and we find the relationship between the two supervision ways. In the next section, we talk about the dataset and processing part of the image: including preprocessing, tumor processing, and CRF as postprocessing. Then we present the details of the implementation of the experiment: details of our model, model pruning method, quality measure, and loss function, which includes: the loss between outputs and labels and the loss between the encoding model and the segmentation model, using a cascaded method to do tumor segmentation. Finally, we give our results of the experiments.

The remainder of the paper is organized as follows. Section \ref{2} describes related works, the most important of which are UNet++(Section \ref{2.3}) and BS-UNet(Section \ref{2.4}). In Section \ref{3}, we propose our network: Encoding feature supervised U-net++(ES-UNet++) by finding the relationship between deep supervision and bottleneck feature supervision and introducing the Network Architecture. Section \ref{4} is the experiment and the results: Dataset and Processing(Section \ref{4.1}); Implementation Details(Section \ref{4.2}); Results(Section \ref{4.3}). Section \ref{5} provides a brief conclusion.

\section{Related works}\label{2}

\subsection{FCN \& U-Net}\label{2.1}

U-Net is actually a kind of fully convolutional network (FCN) that is completely symmetrical in structure without fully connected layers. It is always constructed by an encoder part on the left and a decoder part on the right, which has used a fully convolutional network to address the problem of liver segmentation and detection of liver metastases, which processed whole images rather than patches with a small-scale data set \cite{b1}. V-Net is a fully convolutional end-to-end network depending on volumetrics to measure MRI volumes and predict segmentation of the entire object \cite{b2}. And Christ et al. \cite{b3} presented a fully convolutional cascade network that cascades two FCNs to achieve combined and automatic liver and lesion segmentation. In terms of U-Net, Ronneberger et al. \cite{b4} proposed a network that had a symmetric architecture to segment images in 2D, while 3D U-Net \cite{b5} was proposed for volumetric segmentation of volumetric input images, which shared a structure similar to 2D U-Net but improved the structure by applying 3D operations instead of 2D. In addition to this, FU-Net \cite{b6} presented a feedback-weighted U-Net, based on general U-Net, selecting a dynamically weighted cross-entropy as a loss function. Transformer-UNet \cite{b7} combined transformer and U-Net by adding transformer modules in raw images instead of feature maps in U-Net. 

\subsection{Others Network}\label{2.2}

There are also many networks designed for medical image segmentation. Lu et al. \cite{b8} presented a model for performing automatic liver segmentation. They first used three-dimensional CNN to detect the liver and achieve probability segmentation at the same time. Then, the accuracy of the initial segmentation is refined by the cutting of the graph and the previously learned probability graph. Similarly, Hu et al. \cite{b9} also presented an automatic segmentation network on top of 3D CNN and global optimization of surface evolution. Rezaei et al. \cite{b10} proposed a method for automatic instance semantic segmentation based on DNN, which was customized to represent tumors on MRI and CT images. A multitask learning framework that included three steps was presented, including detection, segmentation, classification, and a new tumor detection technique. LT-Net \cite{b11} presented a one-shot segmentation model via the traditional atlas-based segmentation method, which can release the pressure and heavy work of medical image annotation , and Wang et al. selected forward-backward consistency to conquer the problem of the need for high-level supervision.

\subsection{UNet++}\label{2.3}

Like U-Net, UNet++ begins with an encoding path and a decoding path, but is connected through a redesigned skip connection \cite{b12},\cite{b13}. In this skip connection, the encoder feature maps pass through dense convolution blocks whose number of convolution layers depends on the pyramid level, whereas in U-Net they are concatenated simply to the decoder. The difference in skip connection between UNet++ and U-Net is intended to reduce the semantic gap between the encoder and decoder feature maps by means of dense convolution blocks before fusion. The dense convolution block serves to bring the semantic aspect of the encoder features closer to the corresponding decoder feature maps, since it is assumed that the optimization problem is simpler when they are semantically similar. In addition, deep supervision is used in UNet++ to achieve model pruning and speed gain by selecting a segmentation map from a segmentation branch.
\subsection{Bottleneck feature Supervision U-Net}\label{2.4}

The bottleneck feature Supervision U-Net (BS-UNet) consists of two different U-Net structures, which are the encoding U-Net and the segmentation U-Net \cite{b14}. The main idea of BS U-Net is to provide additional supervision of the bottleneck layer. The encoding U-Net without skip connections can be seen as an auto-encoder with the role of further extracting information from the bottleneck layer, which can subsequently be used to train the segmentation U-Net by minimizing the MSE loss between the two bottleneck feature vectors generated by the encoding U-Net and the segmentation U-Net. After the training of the encoding U-Net, the loss function of the segmentation U-Net will contain two components, the dice loss between the network output and the label map, and the MSE loss between the two bottleneck feature vectors of the encoding U-Net and the segmentation U-Net. According to previous experiments, the supervision of bottleneck information helps speed up training and improve segmentation performance.
\section{Proposed Network Architecture: Encoding feature supervised U-net++(ES-UNet++)}\label{3}

\subsection{Relationship between Deep Supervision and Bottleneck Feature Supervision}\label{3.1}

In this section, we want to adopt bottleneck feature supervision to achieve a similar effect of deep supervision in UNet++.

\subsubsection{Compare Two Networks}\label{3.1.1}

\begin{figure}[htbp]
\centerline{\includegraphics[width=0.7\linewidth]{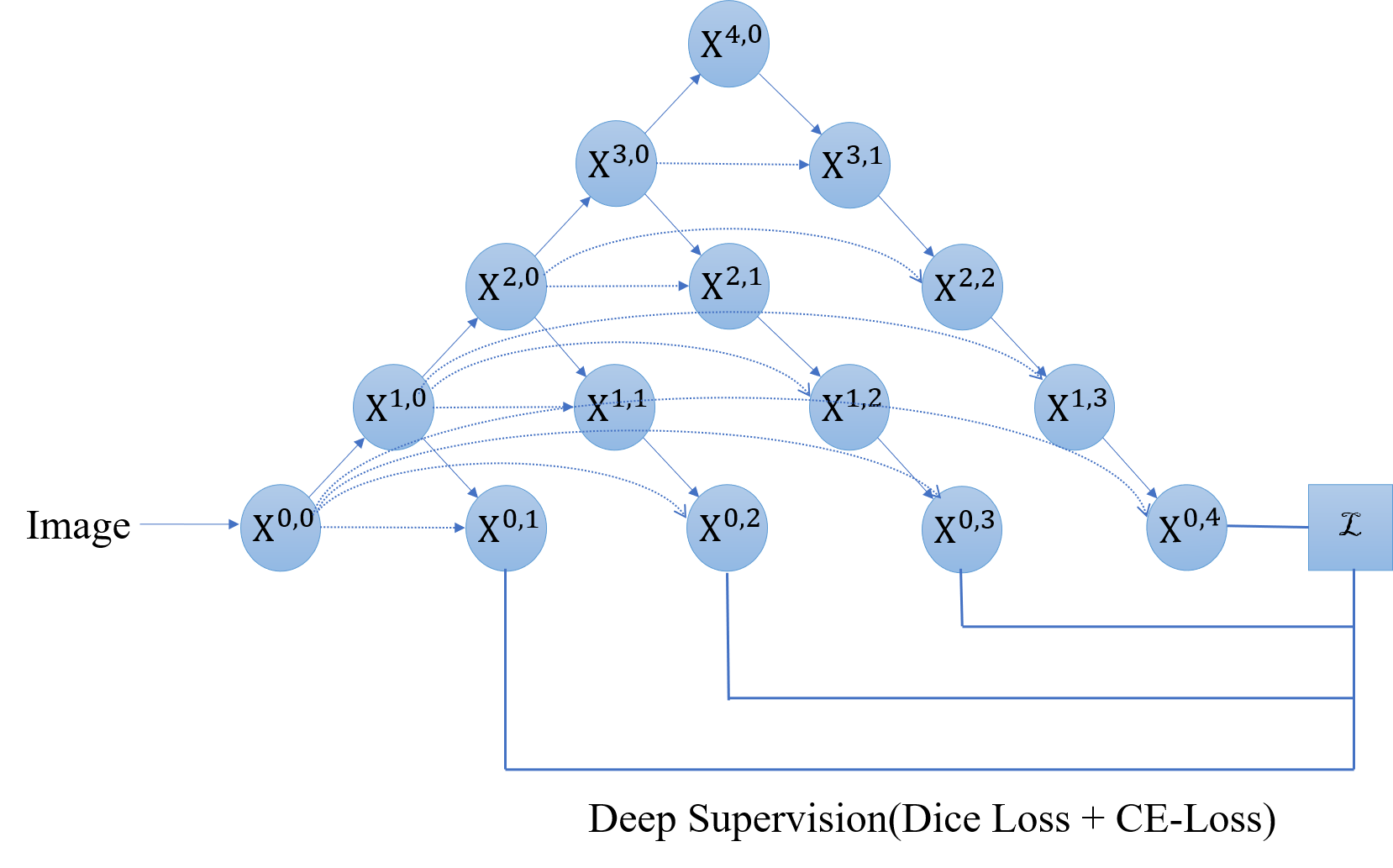}}
\caption{U-Net$^e$}
\label{fig_1}
\end{figure}

U-Net$^e$ is a previous version of UNet++, which combines U-Nets of different depths sharing the same encoder into a unified architecture, as in ``Fig.~\ref{fig_1}''. Compared to UNet++, U-Net$^e$ remains the original skip connections without transforming them into densely connected skip connections, which allows dense feature propagation along skip connections. Under this structure, the shallow decoders $\mathrm{X}^{0,1}$, $\mathrm{X}^{0,2}$, $\mathrm{X}^{0,3}$ cannot be trained in the network, if the deep supervision in U-Net$^e$ is removed.
 
\begin{figure}[htbp]
\centerline{\includegraphics[width=0.7\linewidth]{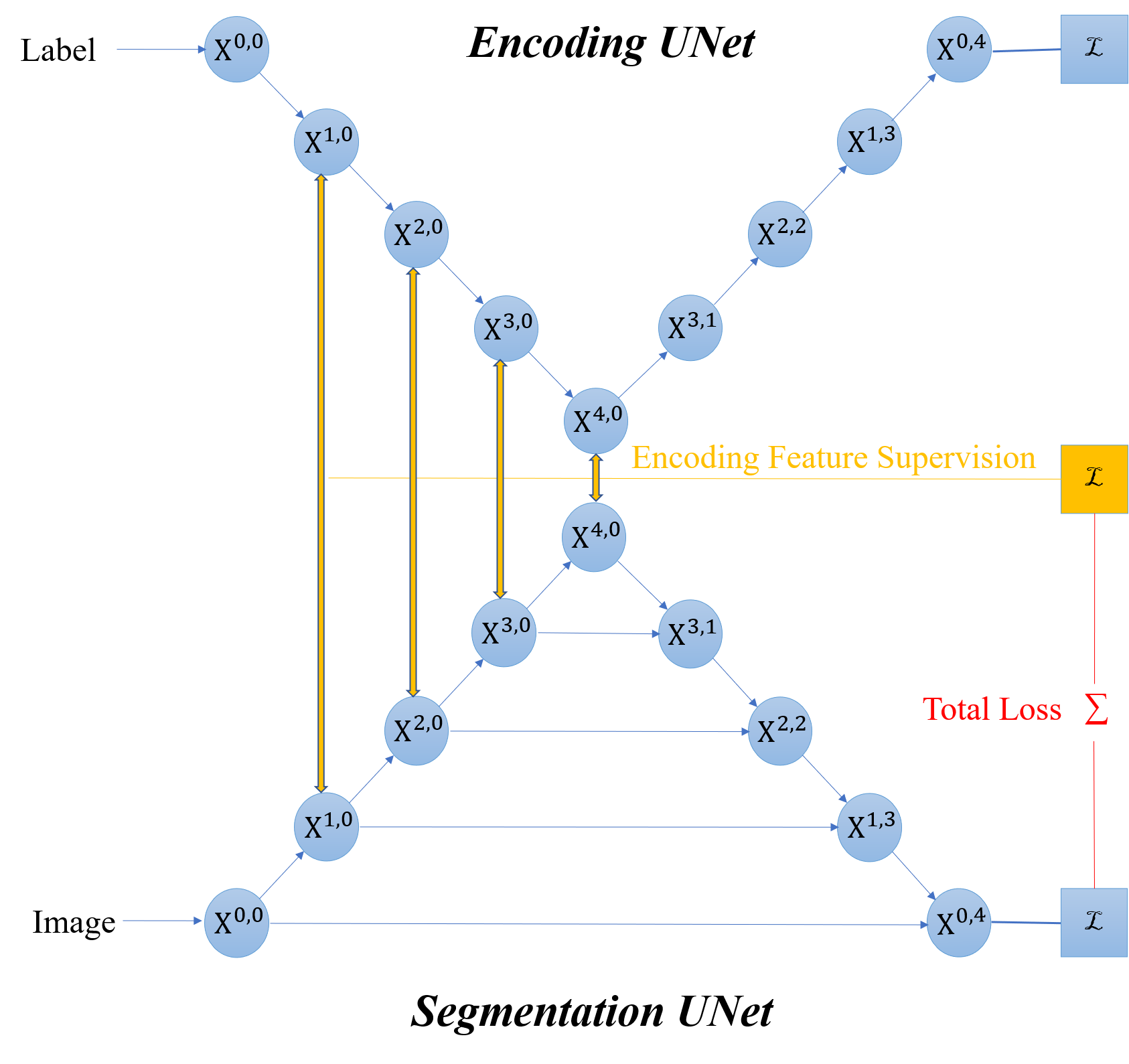}}
\caption{ES-U-Net}
\label{fig_2}
\end{figure}

Inspired by the bottleneck feature supervision, we designed a network based on BS U-Net where we apply the feature supervision to the entire encoding pathway, including the bottleneck as presented in ``Fig.~\ref{fig_2}'', which is supposed to have a similar effect of deep supervision in U-Net$^e$. Unlike U-Net$^e$, the new network performs feature supervision by minimizing the MSE loss between the two feature vectors generated by encoding U-Net and segmentation U-Net before upsampling at each encoding layer. Therefore, we name this new network Encoding Feature Supervised U-Net (ES-U-Net). 

From the experiment results, both U-Net$^e$ and ES-U-Net have segmentation performance similar to basic U-Net, and the training speed of ES-U-Net is better than that of U-Net$^e$ and U-Net.

\subsubsection{Results}\label{3.1.2}

\begin{figure}[htbp]
\centerline{\includegraphics[width=0.9\linewidth]{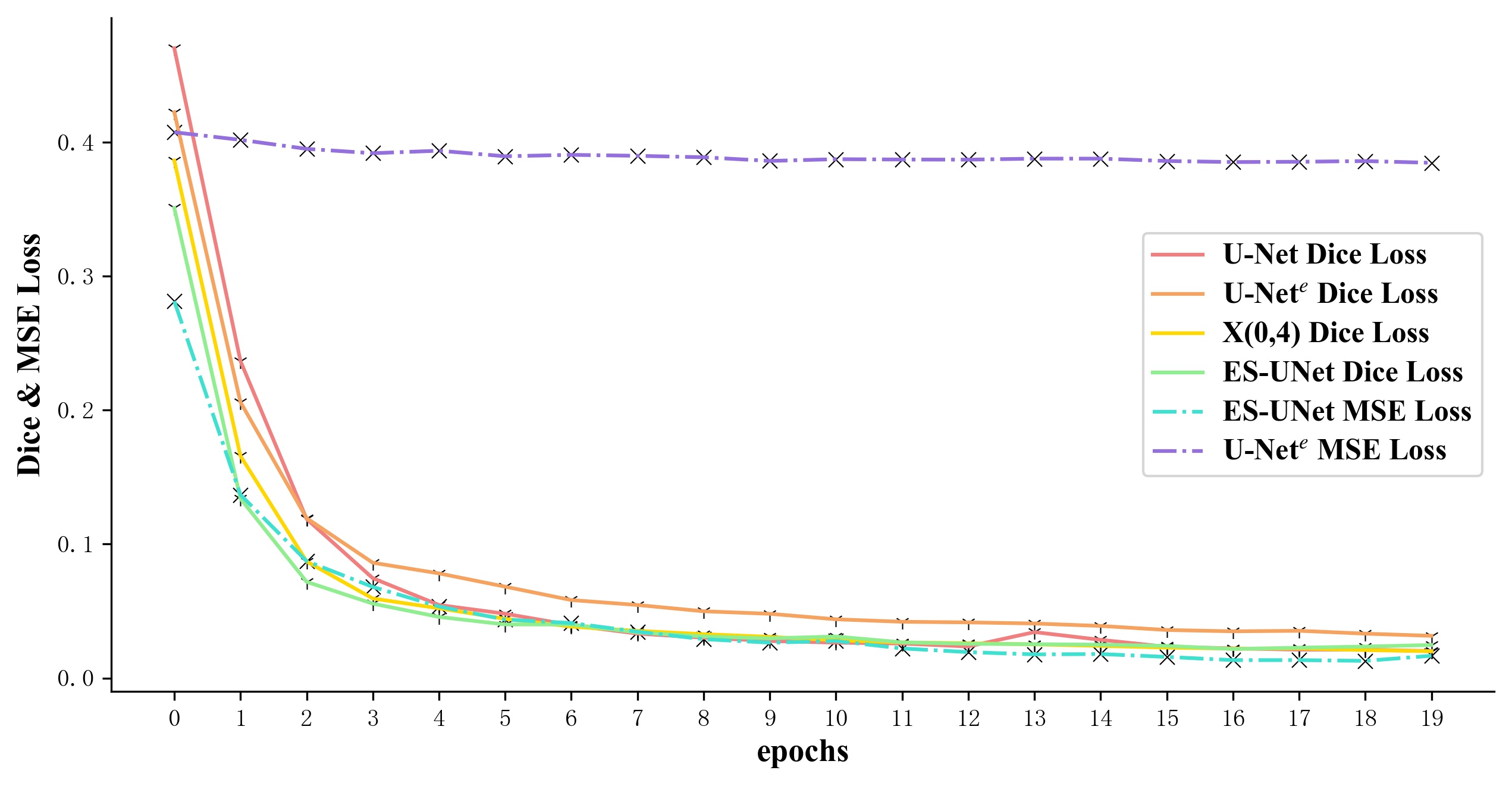}}
\caption{Training Loss}
\label{fig_3}
\end{figure}

We use dice score, dice loss, and MSE loss (Euclidean loss) to evaluate the performance of the three networks in part of our data set. MSE loss is used to measure the difference in encoding features between the network and the encoding U-Net in ES-U-Net. The final dice scores for U-Net, U-Net$^e$ and ES-U-Net on liver segmentation performance are $96.2\%$,$96.0\%$ and $95.3\%$, respectively, and the convergence of dice loss and MSE loss for various versions of U-Net are shown in ``Fig.~\ref{fig_3}''. Based on the results, we find that the three versions of U-Net perform similarly in liver segmentation and all obtain relatively high Dice scores. However, from ``Fig.~\ref{fig_3}'', it can be discovered that ES-U-Net has a higher convergence rate than U-Net and U-Net$^e$, while the MSE losses of ES-U-Net and U-Net$^e$ also show significant differences(in order to draw six loss values in a figure, we halve the MSE loss of U-Net$^e$). The MSE loss of ES-U-Net decreases with dice loss, while the MSE loss of U-Net$^e$ does not show a significant decreasing trend. It indicates that although the ideas of ES-U-Net and U-Net$^e$ are similar, designed to utilize information in various encoding layers, U-Net$^e$ does not have the effect of reducing the average MSE loss between the feature vectors in the U-Net encoding pathways$^e$ and encoding U-Net in ES-U-Net as ES-U-Net does. We hypothesize that the constant MSE loss of U-Net$^e$ is ascribed to the structure of the network. Since we do deep supervision on four decoder outputs: $\mathrm{X}^{0,1}$, $\mathrm{X}^{0,2}$, $\mathrm{X}^{0,3}$, $\mathrm{X}^{0,4}$, the gradients in the encoding features are considerably small during the backpropagation process and therefore it is difficult to train the encoder perfectly. While deep supervision in U-Net$^e$ also has the advantage of its ability to train shallow decoding pathways in the network in addition to the original encoders and decoders by including the output of shallow decoders in the supervision.

\subsection{Our network}\label{3.2}
Based on UNet++ and what we find in the previous section, we propose encoding feature supervised UNet++(ES-UNet++) (``Fig.~\ref{fig_3}''). ES-UNet++ consists of two parts: Encoding UNet++ and Segmentation UNet++.

\begin{figure}[htbp]
\centerline{\includegraphics[width=0.7\linewidth]{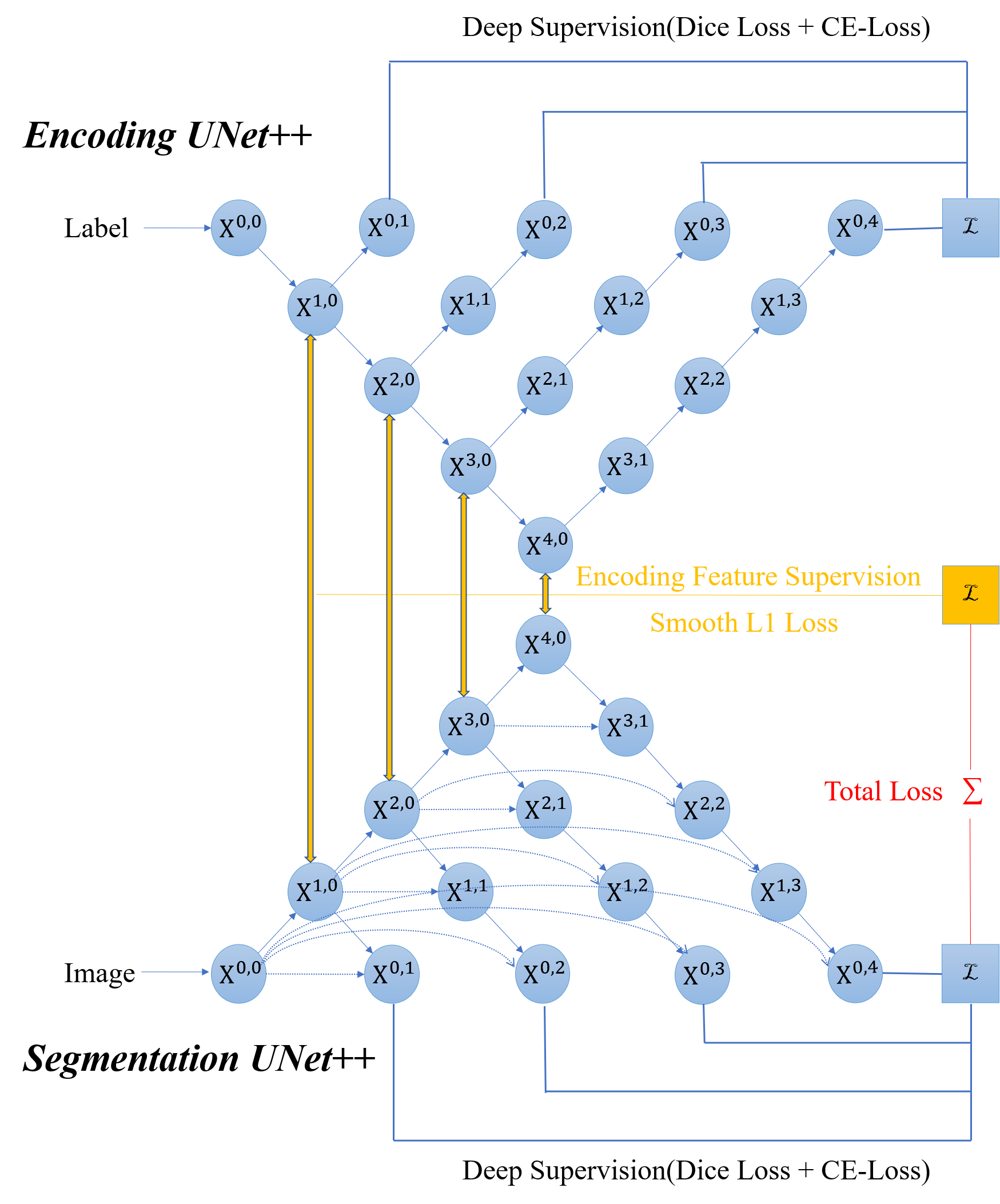}}
\caption{ES-UNet++}
\label{fig_4}
\end{figure}

\subsubsection{Encoding UNet++}\label{3.2.1}

For Encoding UNet++, we have already mentioned the benefits of using this approach. In designing this network structure, we want the two parts of the network to be as similar as possible. So even though we are just using the encoding feature to supervise, we are still using UNet++ as the basis for Encoding UNet++, instead of the U-Net. In addition, we will remove the skip connection because the input and output are almost the same and in this case the feature fusion between different layers has little effect. Like U-Net$^e$, the network with this architecture will not be able to train the features of the middle triangle area, therefore, we use deep supervision.

\subsubsection{Segmentation UNet++}\label{3.2.2}

For segmentation UNet++, it adds supervision of the encoder to basic UNet++. More specifically, it adds Smooth L1 loss between each encoding feature of encoding UNet++ and segmentation UNet++ to the loss function. For this supervision, we have two options. The first is to use Smooth L1 loss as supervision only for the bottleneck features. The advantage of this approach is that it can effectively reduce computational effort because the number of bottleneck feature parameters is minimal. The other option is to supervise each layer of the encoder with Smooth L1 loss. As mentioned earlier, this approach has the similar effect of deep supervision. Furthermore, each encoder unit can be considered as a UNet++ bottleneck of different sizes, leading to a significant improvement in shallow output: $\mathrm{X}^{0,1}$, $\mathrm{X}^{0,2}$, $\mathrm{X}^{0,3}$.

\subsubsection{Training}\label{3.2.3}

Training the ES-UNet++ will follow these steps: (1) Training Encoding UNet++: the labels of the images are used as both the input and the labels to train Encoding UNet++. (2) The original image is inputted to the well-trained Encoding UNet++, and the corresponding image encoding feature is obtained by prediction and saved. (3) Training Segmentation UNet++: It is almost identical to the original UNet++ training method, 
except that the Smooth L1 loss between the encoding feature of two partial UNet++ is added to the overall loss function in ES-UNet++. More details will be mentioned in Section \ref{4.2}.

\section{Experiment and Results}\label{4}

\subsection{Datasets: LiTS \& Pre-Processing \& Post-processing(CRF)}\label{4.1}

\subsubsection{Dataset: LiTS}\label{4.1.1}

LiTS dataset contains 201 contrast-enhanced computed tomography images in DICOM format. Among them, 131 images have been classified as training images and the other 69 images as test images.

The data(or, images) in LiTS are saved in NII form. An NII file expresses 3D information, whereas common image formats (e.g. PNG, JPEG, etc.) can merely express 2D information. From our perspective, an NII file can be seen as a collection of slices. And each slice can be viewed as a common 2D picture.

\subsubsection{Pre-Processing}\label{4.1.2}

Medical images help medical discovery and quantitative analysis of organs of interest, and the U-Net architecture has shown the most advanced performance in medical image segmentation. However, it is still a challenging task to automatically segment the liver from CT images because of its similar pixel values to other adjacent organs in the abdomen, low contrast, uneven intensity, and associated noise. To better classify the liver and its tumor, we need to use image pre-processing to achieve the effect of image enhancement and information extraction.

HU windowing is considered the most significant pretreatment technique because multiple organs have intensity values similar to those of the liver, which makes model learning difficult \cite{b15}. Researchers tend to limit the range of HU values to the target organ. To study the liver, HU should be limited to [-100,400]. 

Medical images obtained by different methods will also suffer from poor contrast, which will lead to image quality degradation. The application of adaptive histogram equalization technology can improve image quality and obtain a better segmentation effect later \cite{b16}. Adaptive histogram equalization is used in many image processing studies but is not effective in CT images. Therefore, CLAHE is employed in this study to overcome the issue of overamplification. In all experiments, CLAHE implements parameters 4, clip limit 4, and tile size (8,8).

At the same time, we also normalize the data, which are usually used to replace the range of pixel intensity values with a common scale, adjust the scope of the training set according to the label, and optimize the spatial structure of the NIFTI data to save storage space. After processing, the data are fed into a learning model for automatic liver segmentation(from ``Fig.~\ref{fig_5}'' to ``Fig.~\ref{fig_6}'').

\subsubsection{Tumor-Processing}\label{4.1.3}

\begin{figure}
  \centering
  \begin{minipage}[t]{0.22\linewidth}
    \centering
    \centerline{\includegraphics[width=0.9\linewidth]{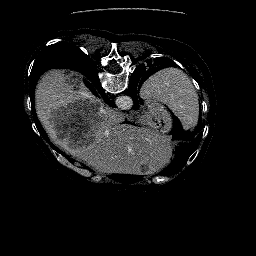}}
    \caption{original}
    \label{fig_5}
  \end{minipage}
  \begin{minipage}[t]{0.22\linewidth}
    \centering
    \centerline{\includegraphics[width=0.9\linewidth]{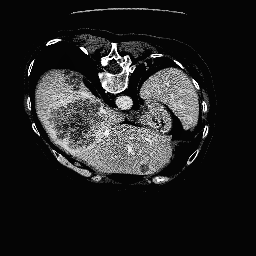}}
    \caption{process}
    \label{fig_6}
  \end{minipage}
  \begin{minipage}[t]{0.22\linewidth}
    \centering
    \centerline{\includegraphics[width=0.9\linewidth]{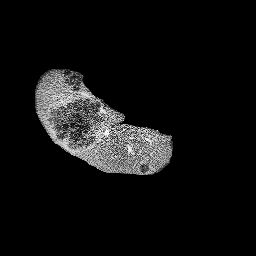}}
    \caption{liver}
    \label{fig_7}
  \end{minipage}
  \begin{minipage}[t]{0.22\linewidth}
    \centering
    \centerline{\includegraphics[width=0.9\linewidth]{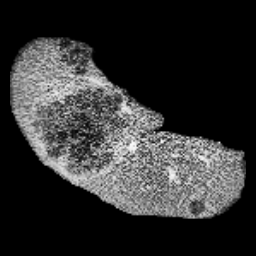}}
    \caption{resized}
    \label{fig_8}
  \end{minipage}
\end{figure}

Direct tumor segmentation is not effective, so we adjusted the method of tumor segmentation. First, we extracted the results of the successfully segmentation liver as the source of tumor segmentation instead of the original image(from ``Fig.~\ref{fig_6}'' to ``Fig.~\ref{fig_7}''). Then the effective part of the image, the liver part after the first segmentation, is extracted and normalized, which improved the training effect (``Fig.~\ref{fig_8}'') \cite{b14}.

\subsubsection{Post-Processing(CRF)}\label{4.1.4}

After the neural network, Fully Connected CRFs can be used to solve the detail segmentation problem, which is independent of each other\cite{b17},\cite{b18}. For each segmentation image of $N$ pixels, denoted by $\mathbf{I}=\left\{I_{1}, I_{2}, \cdots I_{N}\right\}$, and a set of category labels $\mathcal{L}=\left\{l_{1}, l_{2}, \cdots, l_{k}\right\}$, in this article $k=2$, since there are two categories (when performing liver segmentation: liver and background; when performing cascaded tumor segmentation: tumor and liver), and labels are denoted by $\mathbf{X}=\left\{X_{1}, X_{2}, \cdots X_{N}\right\}$. For conditional random fields $(\mathbf{X}, \mathbf{I})$, it is characterized by a Gibbs distribution:

\begin{equation}
P(\mathbf{X} \mid \mathbf{I})=\frac{1}{Z(\mathbf{I})} \exp (-E(\mathbf{X}, \mathbf{I}))
\end{equation}

where $Z$ is the normalization coefficient and $E(\mathbf{X}, \mathbf{I})$ represents the label distribution and the Gibbs energy function is defined as:

\begin{equation}
Z(\mathbf{I})=\sum_{x} \exp (-E(\mathbf{X}, \mathbf{I}))
\end{equation}

\begin{equation}
E(\mathbf{x} \mid \mathbf{I})=\sum_{i} \psi_{u}\left(x_{i} \mid \mathbf{I}\right)+\sum_{i<j} \psi_{p}\left(x_{i}, x_{j} \mid \mathbf{I}\right)
\end{equation}

$\psi_{u}$ is a one-dimensional potential function that is calculated by the neural network: ES-UNet++. And $\psi_{p}$ is a binary potential function, which is defined as:

\begin{equation}
\psi_{p}\left(x_{i}, x_{j} \mid \mathbf{I}\right)=\mu\left(x_{i}, x_{j}\right) \underbrace{\sum_{m=1}^{K} w^{(m)} k^{(m)}\left(\mathbf{f}_{i}, \mathbf{f}_{j}\right)}_{k\left(\mathbf{f}_{i}, \mathbf{f}_{j}\right)}
\end{equation}

where $k(\mathbf{f} * i, \mathbf{f} * j)$ denotes:

\begin{equation}
k(\mathbf{f} * i, \mathbf{f} * j) = w^{(1)} \exp \left(-\frac{\left|p_{i}-p_{j}\right|^{2}}{2 \theta_{\alpha}^{2}}-\frac{\left|I_{i}-I_{j}\right|^{2}}{2 \theta_{\beta}^{2}}\right) + w^{(2)} \exp \left(-\frac{\left|p_{i}-p_{j}\right|^{2}}{2 \theta_{\gamma}^{2}}\right)
\end{equation}

CRFs aim to assign each pixel of an image by solving the maximum posterior probability(MAP):

\begin{equation}
x^{*}=\arg \max _{x \in L^{N}} P(x \mid I)
\end{equation}

Thus, pixels with large similarities are labeled as the same category, and different category labels are given to pixels with small similarities.

\subsection{Implementation Details}\label{4.2}

\subsubsection{Details of ES-UNet++}\label{4.2.1}

There are three operations in the entire network: downsampling, upsampling, and skip connection. More details are given in the ``TABLE.~\ref{tab_1}''.

\begin{itemize}
\item downsampling: The downsampling consists of two convolution blocks, the first one(Downsampling\_1) doubles the number of channels and the second one(Downsampling\_2) remains unchanged. Each module consists of a convolution layer, a batch normalization layer, and an activation function(ReLu).
\item upsampling: The upsampling goes through a transposed convolution layer first, halving the number of channels and enlarging the feature map, and is initialized in bilinear interpolation. In particular, the transposed convolution in the UNet++ encoder does not change the number of channels because it does not have a skip connection.
\item Skip Connection \& Convolution\_SC: After upsampling, the feature map concatenates the features from the several skip connection. Then, after a convolution block, its depth changes to the appropriate number. The block's convolution layer is called Convolution\_SC and its input channels depend on the upsampling feature map and several skip connection feature maps. Its output channel depends on the number of the channel of the encoding feature map, which is at the same level of the layer.  
\end{itemize}

\begin{table}[htbp]
\caption{Details of modules in the ES-UNet++}
\begin{center}
\begin{tabular}{|c|c|c|c|c|c|c|}
\hline
                 & In & Out & type        & Kernel & Stride & Pad \\ \hline
Convolution\_1   & 2n & n   & Convolution & (3,3)  & 1      & 1       \\ \hline
Convolution\_2   & n  & n   & Convolution & (3,3)  & 1      & 1       \\ \hline
Max pooling      & n  & n   & pooling     & (2,2)  & 2      & 0       \\ \hline
upsampling      & n  & 2n  & Trans-Conv  & (4,4)  & 2      & 1       \\ \hline
Convolution\_SC  & x  & y   & Convolution & (3,3)  & 1      & 1       \\ \hline
\multicolumn{4}{l}{In: input channel \quad Out: output channel}  
\end{tabular}
\label{tab_1}
\end{center}
\end{table}

\subsubsection{Model Pruning}\label{4.2.2}

Instead of using $\mathrm{X}^{0,4}$ as the output, we can choose $\mathrm{X}^{0,1}$, $\mathrm{X}^{0,2}$, $\mathrm{X}^{0,3}$. In this way, it prunes UNet++ into smaller UNet++, which can reduce inference time \cite{b12}. Reference [12] shows that the inference time of UNet++ is about: 16s ($\mathrm{X}^{0,1}$), 23s ($\mathrm{X}^{0,2}$), 37s ($\mathrm{X}^{0,3}$), 55s ($\mathrm{X}^{0,4}$).

\subsubsection{Quality Measure}\label{4.2.3}

We use Dice per case (DPC) and Dice global (DG) as our main evaluation metrics. Since these metrics are only applicable to binary segmentation, we use them in liver and tumor segmentation, respectively. In both segments, we set the pixels within the target object to be 1, representing the foreground, and set the rest to be 0, representing the background. The ground-truth binary volumes are represented by $GT$, the binary volume of model prediction is represented by $MP$, and the Dice score is computed as follows:

\begin{equation}
    D I C E(G T, M P)=\frac{2|G T \cap M P|}{|G T|+|M P|}
\end{equation}

where the Dice score range is $[0,1]$, a flawless segmentation produces a Dice score of 1. A global Dice score(DG) is applied across all cases within the entire volume, while a Dice per case (DPC) is applied per case and is averaged over all cases.

\subsubsection{Loss} \label{4.2.4}

There are six loss functions in our article: dice loss, weighted dice loss, cross-entropy loss, focal loss and segmentation loss between the outputs($\mathrm{X}^{0,1}$, $\mathrm{X}^{0,2}$, $\mathrm{X}^{0,3}$, $\mathrm{X}^{0,4}$) and labels. Smooth L1 loss is the encoding feature map($\mathrm{X}^{1,0}$, $\mathrm{X}^{2,0}$, $\mathrm{X}^{3,0}$, $\mathrm{X}^{4,0}$) between Encoding UNet++ and Segmentation UNet++.

\begin{itemize}
\item Dice Loss(DS-Loss) \& Weighted Dice Loss: First, we introduce the dice loss, which is based on the function of the dice score. According to \cite{b2}, dice loss is widely used in medical image segmentation, since it aims to alleviate the problem of data imbalance. Given the ground truth($GT$) and the model prediction($MP$), the dice loss between $GT$ and $MP$ is calculated as follows:

\begin{equation}
    \mbox{Dice Loss} =1-\frac{2|GT \bigcap MP|}{|GT|+|MP|}
\end{equation}

Then the liver segmentation performance of the liver boundary is relatively poorer than that of the internal regions due to the generally low intensity of the tumor, the liver boundary, and the adjacent regions. To distinguish these regions, we need to make the neural network pay more attention to the boundary regions, that is, to compute a weight map of the same size. We use the method to calculate the weight map in \cite{b14}. Before obtaining the weight map, we first need to calculate a distance map $D$, which is the distance from one pixel to the nearest pixel on the liver boundary. The weight map $W$ can then be computed as follows:
\begin{equation}
A=(w \times F+1) e^{-\frac{D}{2 \sigma^{2}}} \\
\end{equation}
\begin{equation}
W=\frac{A-\min A}{\max A-\min A} 
\end{equation}

where $F$ is the binary (0-1) matrix of the region-of-interest (ROI). $w$ represents the level of importance of the region-of-interest represented by $F$, and $\sigma$ represents the variance. In our experiments, $w$ and $\sigma$ are empirically chosen to be 0.05 and 20, respectively.
By adding the weight map $W$(``Fig.~\ref{fig_10}'') to the dice loss function, the weighted dice loss between $GT$ and $MP$ can be represented by:
\begin{equation}
\mbox{ Weighted Dice-Loss }=1-\frac{2|W \times GT \cap MP|}{(|W \times GT|+|W \times MP|)}
\end{equation}

\begin{figure}[htbp]
\centering
  \begin{minipage}[t]{0.4\linewidth}
    \centering
    \centerline{\includegraphics[width=0.9\linewidth]{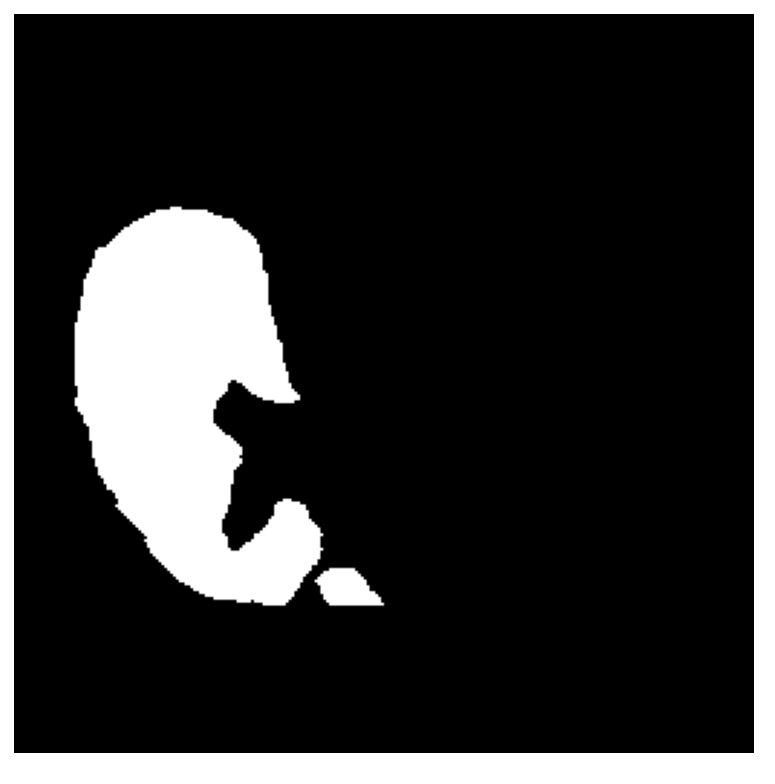}}
    \caption{original}
    \label{fig_9}
  \end{minipage}
  \begin{minipage}[t]{0.4\linewidth}
    \centering
    \centerline{\includegraphics[width=1.08\linewidth]{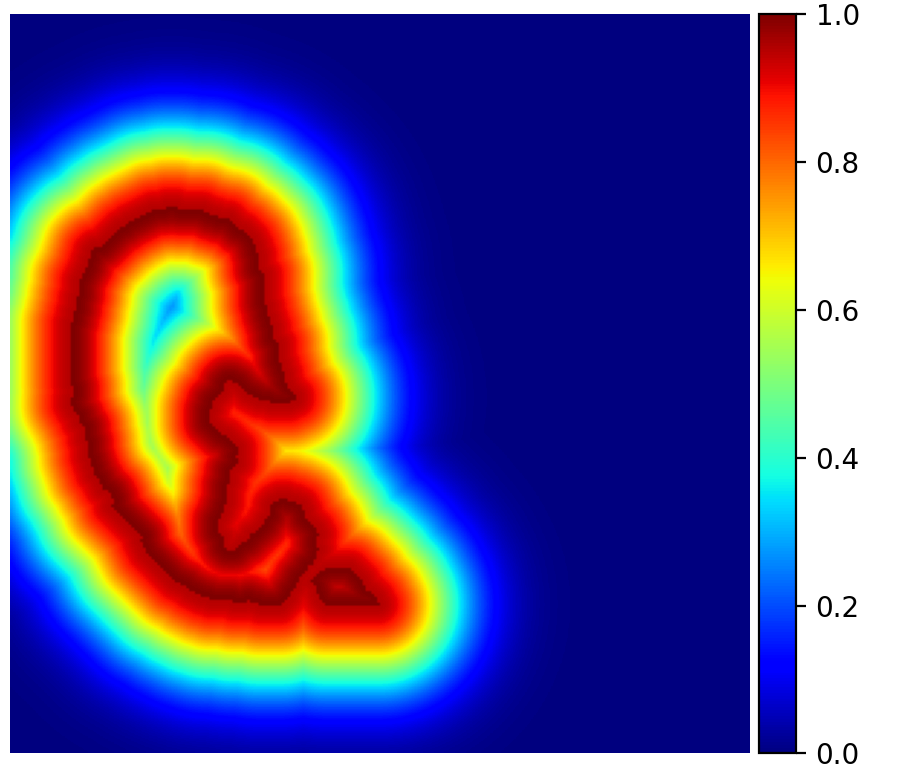}}
    \caption{weight map}
    \label{fig_10}
  \end{minipage}
\end{figure}

\item Cross-Entropy \& Focal-Loss: We use normal cross-entropy loss(CE-Loss) between output and labels to train the network. However, when tumor segmentation is done, the cross-entropy loss will be much smaller than the dice loss after several epochs because the tumor is very small. Therefore, we use the focal loss to handle the class imbalance of pixels\cite{b19}.

\begin{equation}
\mbox{ Focal-Loss }=\left\{\begin{array}{rll}
-(1-p)^{\gamma} \log (p), & \mbox { if } & y=1 \\
-p^{\gamma} \log (1-p), & \mbox { if } & y=0
\end{array}\right.
\end{equation}

\item Segmentation Loss(Seg-Loss): We use the loss function, which combines the cross-entropy loss and the dice loss. Cross-entropy loss has a smooth gradient, and dice loss contributes to class imbalance. 

\begin{equation}
\mbox { Seg-Loss }=w_{1} \times \mbox { Dice Loss }+w_{2} \times \mbox{ CE-Loss }
\end{equation}

\item Deep Supervision: Except $\mathrm{X}^{0,4}$, each $\mathrm{X}^{0,1}$, $\mathrm{X}^{0,2}$, $\mathrm{X}^{0,3}$ is a full-resolution feature
map at multiple semantic levels, which also can be seen as output. Training these can benefit the entire ES-UNet++ and $\mathrm{X}^{0,4}$ \cite{b12}, \cite{b20}.

\begin{equation}
\mbox { DS-Loss }=\sum_{i} w_{i} \times \mbox { Seg-Loss}(\mathrm{X}^{0,i},\mbox {Labels})
\end{equation}

if do not use deep supervision, set $i=4$.

\item Huber Loss \& Smooth L1 Loss:
\begin{equation}
\mbox { Huber Loss }=\sum_{i}l_{i}
\end{equation}
where $l_{i}$ denotes:
\begin{equation}
l_{n}=\left\{\begin{array}{ll}
0.5\left(x_{n}-y_{n}\right)^{2},  & \mbox { if }  \left|x_{n}-y_{n}\right|<\beta \\
\beta *\left(\left|x_{n}-y_{n}\right|-0.5 *\beta\right), & \mbox { otherwise }
\end{array}\right.
\end{equation}
$x_{n}$ is the flattened encoding feature map ($\mathrm{X}^{1,0}$, $\mathrm{X}^{2,0}$, $\mathrm{X}^{3,0}$, $\mathrm{X}^{4,0}$) of Encoding UNet++ and $y_{n}$ is the flattened encoding feature map ($\mathrm{X}^{1,0}$, $\mathrm{X}^{2,0}$, $\mathrm{X}^{3,0}$, $\mathrm{X}^{4,0}$) of Segmentation UNet++. We set $\beta = 1$, then the loss is equivalent to Smooth L1 Loss, which is written as SL1-Loss.

SL1-Loss is less sensitive to outliers than MSE loss. Although we use Batch Normalization, there is still some large number in the feature map after the activation layer(ReLu). And when encoding feature supervision convergences, encoding and segmentation UNet++ become similar in the encoding feature, SL1-Loss changes into L2 Loss which is smoother than L1 Loss so that it is a better optimization process.

\item Total Loss:
\begin{equation}
\mbox { Total loss }=w_{1} \times \mbox { DS-Loss }+w_{2} \times \mbox { SL1-Loss }
\end{equation}
\end{itemize}

In this paper, we set each weighted $w_{i}$ as equally weighted.

\subsubsection{Cascaded}\label{4.2.5}

Instead of using a three-class classification to segment the liver and tumor at one time, we first segment the liver and then segment the tumor within the liver region using another network \cite{b21}. Specifically, we first use ES-UNet++(called liver segmentation ES-UNet++) to segment the tumor of the preprocessed images in LiTS. Then, process the image mentioned in Section \ref{4.1.3}. Using another ES-UNet++(called tumor segmentation ES-UNet++) to segment tumors from liver images. The two ES-UNet++ are cascaded when tumor segmentation is performed. We use transfer learning to train the tumor segmentation ES-UNet++ from the liver one.

\subsection{Results}\label{4.3}

In this section, we compare six forms of ES-UNet++, which is UNet++(UN), bottleneck feature supervision UNet++(BS), encoding feature supervision UNet++(ES) and normal supervision(NS), deep supervision(DS). For example, ES-DS means encoding feature supervision UNet++ and using deep supervision. When we use deep supervision, the dice loss will be the average of four outputs' dice loss. Therefore, we cannot compare the dice loss between DS and NS, but the dice score is available. Additionally, the Smooth L1 loss cannot also be compared between BS and ES.

\subsubsection{Training Dice Loss}\label{4.3.1}

\begin{figure}[htbp]
\centerline{\includegraphics[width=0.9\linewidth]{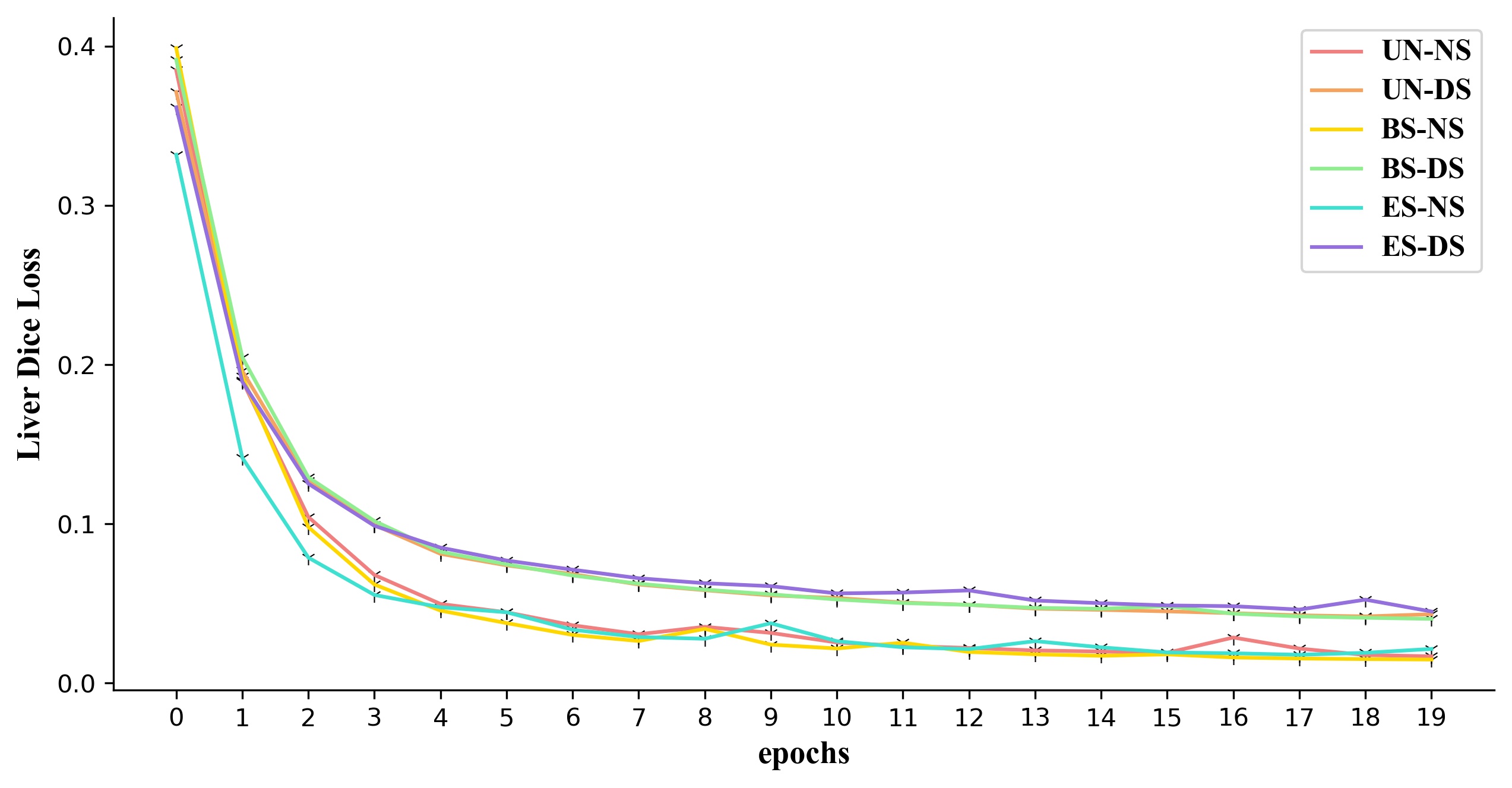}}
\caption{Liver Dice Loss}
\label{fig_11}
\end{figure}

As shown in ``Fig.~\ref{fig_11}'', in liver segmentation, encoding features and bottleneck supervision can foster convergence without deep supervision. However, their effects tend to diminish when deep supervision is deployed because deep supervision affects the encoding path of U-net++, just like encoding features and bottleneck supervision do. 

\begin{figure}[htbp]
\centerline{\includegraphics[width=0.9\linewidth]{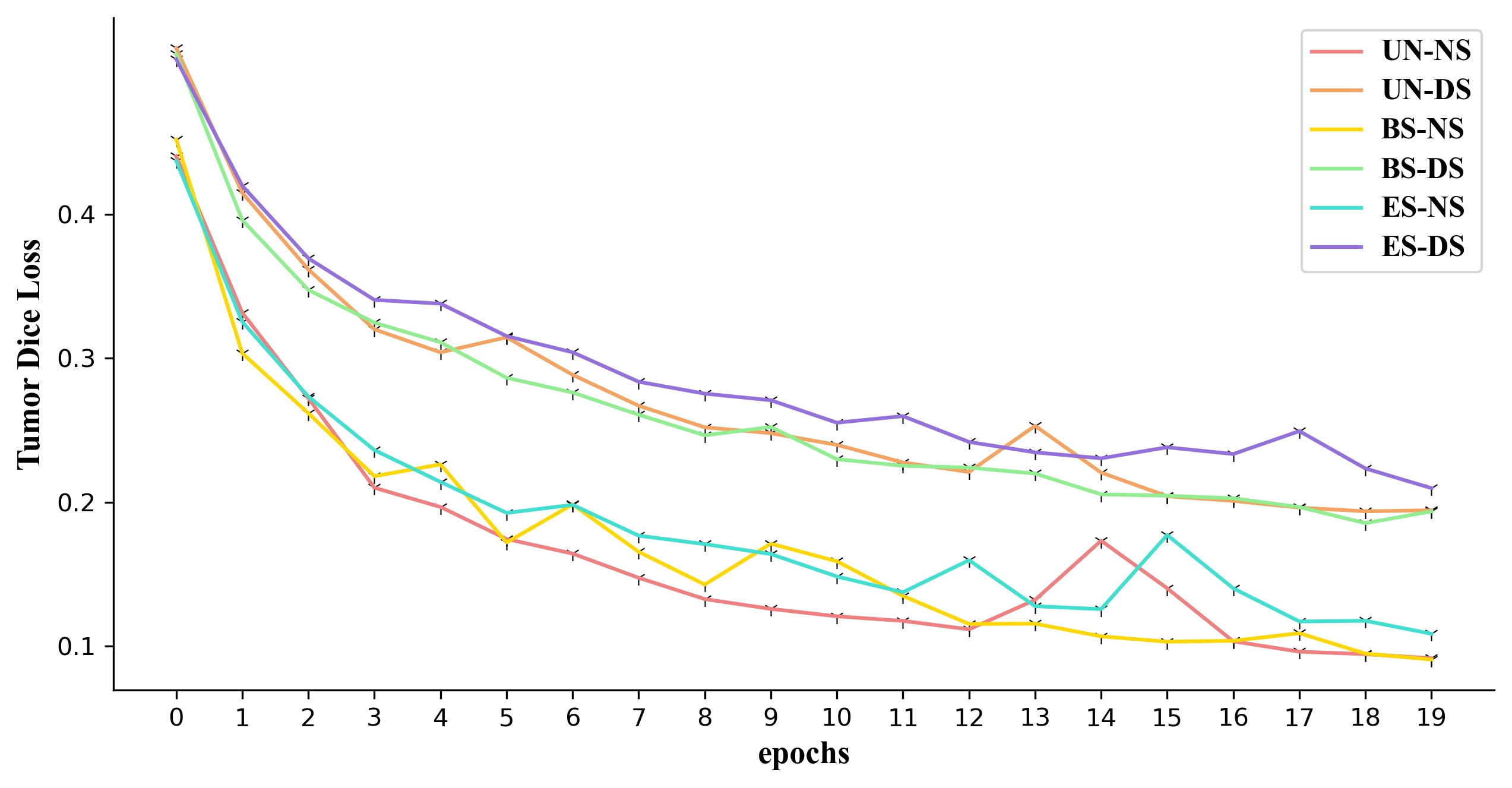}}
\caption{Tumor Dice Loss}
\label{fig_12}
\end{figure}

As shown in ``Fig.~\ref{fig_12}'', in tumor segmentation, bottleneck supervision can accelerate loss decay, but it seems that encoding feature supervision delayed convergence. Maybe that is because $\mathrm{X}^{0,4}$ showed the best performance and the effect on $\mathrm{X}^{0,4}$ of the encoding feature supervision is less than that of the bottleneck supervision.

\subsubsection{Training Smooth L1 Loss}\label{4.3.2}

\begin{figure}[htbp]
\centerline{\includegraphics[width=0.9\linewidth]{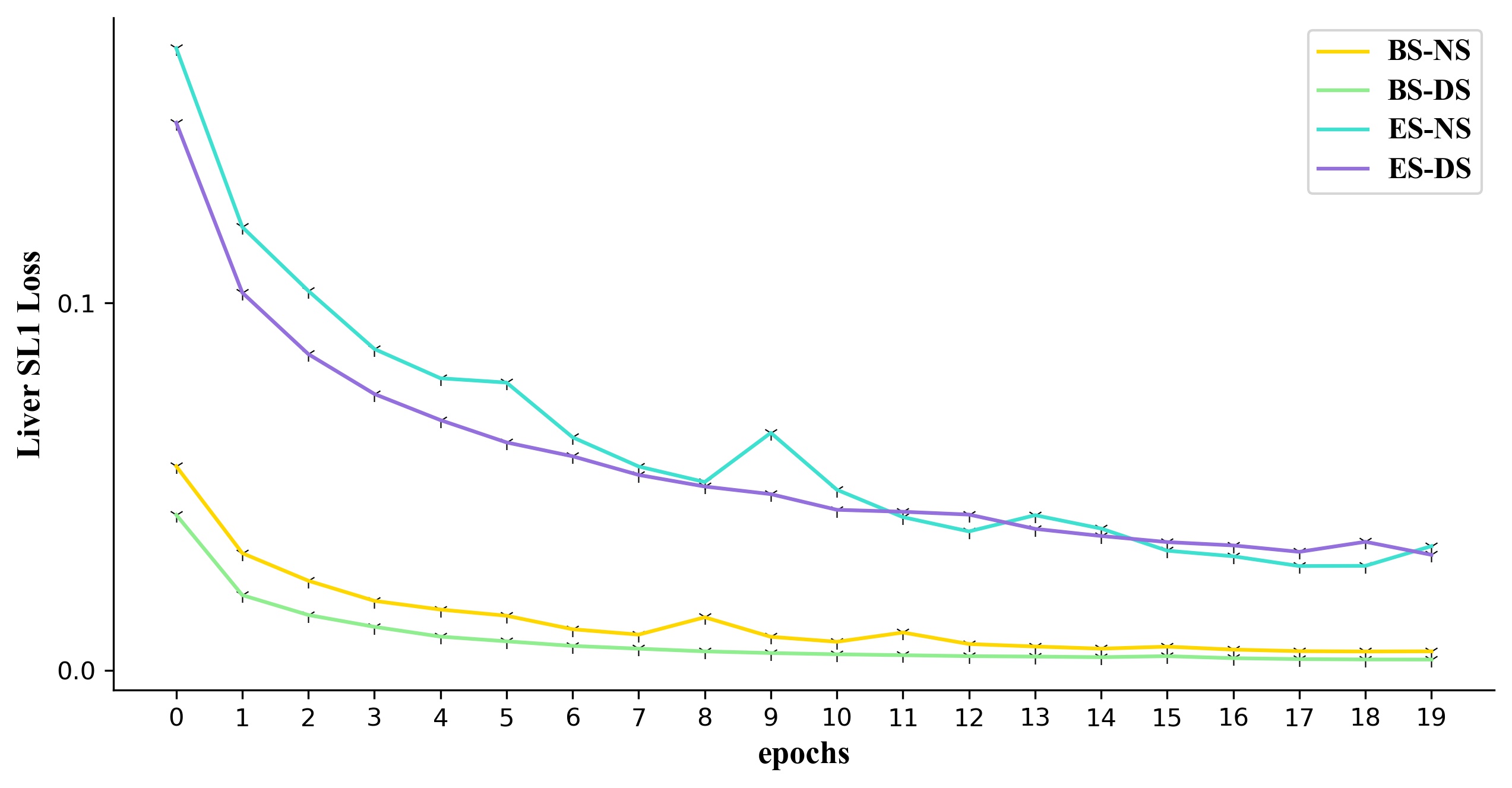}}
\caption{Liver SL1-Loss}
\label{fig_13}
\end{figure}

In liver segmentation, deep supervision can accelerate convergence compared to absence of supervision, because deep supervision has an effect on the encoding path during backpropagation, as seen in ``Fig.~\ref{fig_13}''.

\begin{figure}[htbp]
\centerline{\includegraphics[width=0.9\linewidth]{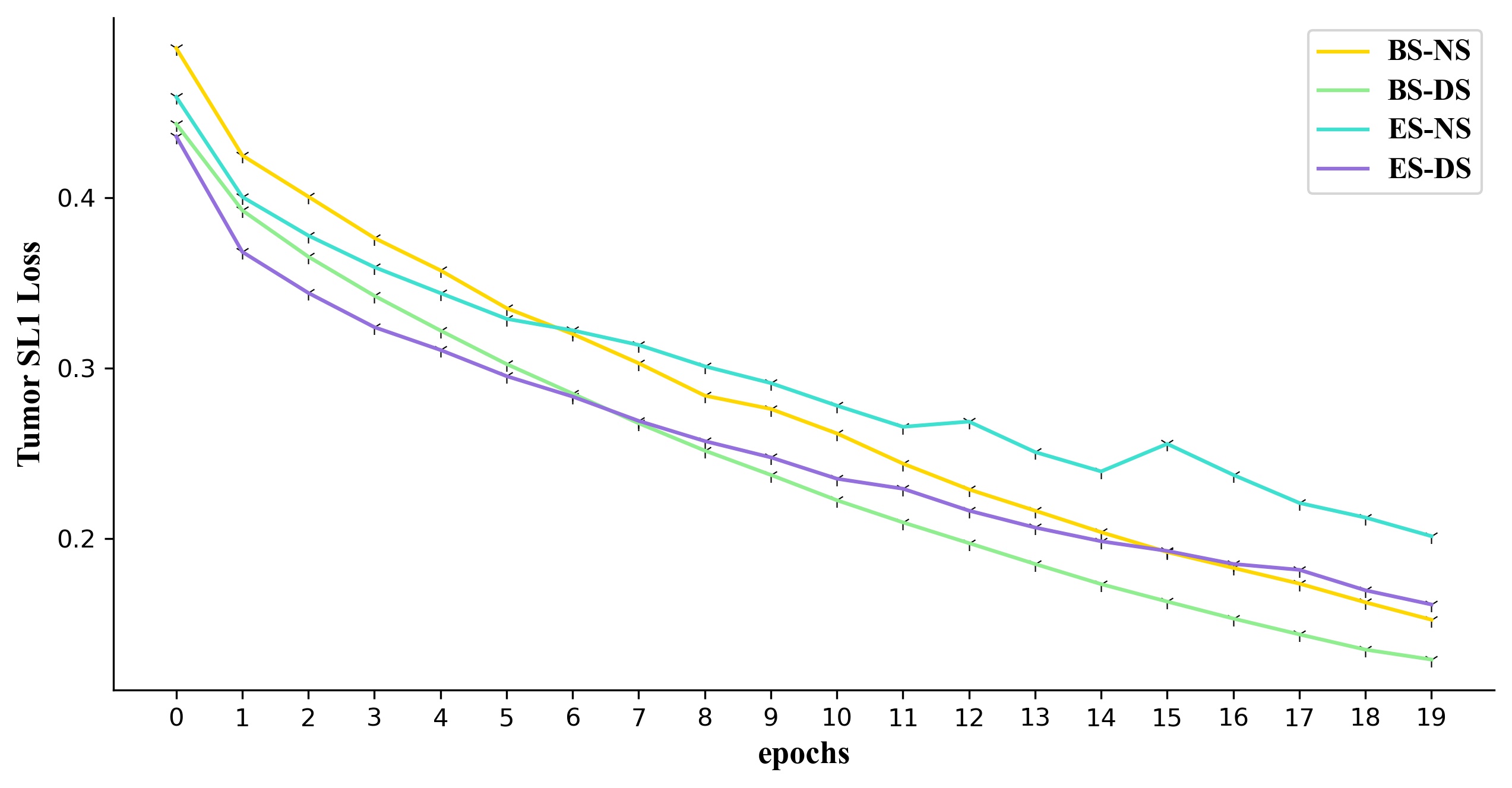}}
\caption{Tumor SL1-Loss}
\label{fig_14}
\end{figure}

In tumor segmentation, deep supervision is performed similarly, as seen in ``Fig.~\ref{fig_14}''. Besides, encoding feature supervision accelerated the convergence at first, because it supervises all 4 feature maps, and so its influence on the encoding path is stronger. However, after several epochs, the L1 losses of encoding features supervision are overtaken by the bottleneck, because $\mathrm{X}^{0,4}$ learn the best and the impact of bottleneck supervision on it is more intense than that of encoding features supervision.

\subsubsection{Dice score}\label{4.3.3}

``TABLE.~\ref{tab_2}'' shows the Dice score on the liver and tumor segmentation test set. We compare six forms of network and their four outputs:$\mathrm{X}^{0,1}$, $\mathrm{X}^{0,2}$, $\mathrm{X}^{0,3}$, $\mathrm{X}^{0,4}$.

In liver segmentation test data, encoding feature supervision can bottleneck supervision can help improve Dice scores, especially on $\mathrm{X}^{0,1}$, $\mathrm{X}^{0,2}$, $\mathrm{X}^{0,3}$. But if deep supervision is disabled, the performance of $\mathrm{X}^{0,1}$, $\mathrm{X}^{0,2}$, $\mathrm{X}^{0,3}$ will not be improved by the supervision of bottleneck or encoding features, because the 2 supervision methods cannot affect the decoding path during backpropagation. 

Regarding tumor segmentation, encoding features and bottleneck supervision can help improve the performance of $\mathrm{X}^{0,1}$, $\mathrm{X}^{0,2}$, $\mathrm{X}^{0,3}$, $\mathrm{X}^{0,4}$, and the effect of encoding features supervision is seemingly stronger than that of bottleneck supervision. Meanwhile, deep supervision can also enhance segmentation.

In Section \ref{4.2.2}, we have mentioned the model pruning. It can shorten the inference time, however, and may reduce the segmentation performance. When the encoding feature supervision is added, the segmentation ability of $\mathrm{X}^{0,1}$, $\mathrm{X}^{0,2}$, $\mathrm{X}^{0,3}$ improves. Therefore, this method can benefit the model pruning to all networks based on UNet++.

\begin{table}[htbp]
\caption{Dice score}
\setlength{\tabcolsep}{1.2mm}{
\begin{center}
\begin{tabular}{|l|llll|llll|}
\hline
\multirow{2}{*}{} & \multicolumn{4}{c|}{Liver Dice score}                                                                                                            & \multicolumn{4}{c|}{Tumor Dice score}                                                                                                            \\ \cline{2-9} 
                  & \multicolumn{1}{l|}{$\mathrm{X}^{0,1}$} & \multicolumn{1}{l|}{$\mathrm{X}^{0,2}$} & \multicolumn{1}{l|}{$\mathrm{X}^{0,3}$} & $\mathrm{X}^{0,4}$ & \multicolumn{1}{l|}{$\mathrm{X}^{0,1}$} & \multicolumn{1}{l|}{$\mathrm{X}^{0,2}$} & \multicolumn{1}{l|}{$\mathrm{X}^{0,3}$} & $\mathrm{X}^{0,4}$ \\ \hline
UN-NS                   & \multicolumn{1}{l|}{1.9\%}                         & \multicolumn{1}{l|}{16.5\%}                        & \multicolumn{1}{l|}{15.7\%}                        & 95.2\%                        & \multicolumn{1}{l|}{20.8\%}             & \multicolumn{1}{l|}{39.0\%}             & \multicolumn{1}{l|}{61.7\%}             & 63.3\%             \\ \hline
UN-DS                   & \multicolumn{1}{l|}{86.4\%}                        & \multicolumn{1}{l|}{91.7\%}                        & \multicolumn{1}{l|}{94.1\%}                        & 95.3\%                        & \multicolumn{1}{l|}{47.1\%}             & \multicolumn{1}{l|}{59.2\%}             & \multicolumn{1}{l|}{63.9\%}             & 65.3\%             \\ \hline
BS-NS                   & \multicolumn{1}{l|}{7.3\%}                         & \multicolumn{1}{l|}{4.2\%}                         & \multicolumn{1}{l|}{8.8\%}                         & 96.1\%                        & \multicolumn{1}{l|}{43.7\%}              & \multicolumn{1}{l|}{49.0\%}             & \multicolumn{1}{l|}{61.7\%}             & 66.5\%             \\ \hline
BS-DS                   & \multicolumn{1}{l|}{87.2\%}                        & \multicolumn{1}{l|}{93.5\%}                        & \multicolumn{1}{l|}{95.5\%}                        & 95.6\%                        & \multicolumn{1}{l|}{59.6\%}             & \multicolumn{1}{l|}{63.2\%}             & \multicolumn{1}{l|}{65.0\%}             & 66.6\%             \\ \hline
ES-NS                   & \multicolumn{1}{l|}{14.0\%}                        & \multicolumn{1}{l|}{21.6\%}                        & \multicolumn{1}{l|}{20.4\%}                        & 95.6\%                        & \multicolumn{1}{l|}{51.6\%}             & \multicolumn{1}{l|}{47.0\%}             & \multicolumn{1}{l|}{63.1\%}             & 66.7\%             \\ \hline
ES-DS                   & \multicolumn{1}{l|}{{ 87.3\%}} & \multicolumn{1}{l|}{{ 92.5\%}} & \multicolumn{1}{l|}{{ 95.0\%}} & { 95.6\%} & \multicolumn{1}{l|}{60.5\%}             & \multicolumn{1}{l|}{63.7\%}             & \multicolumn{1}{l|}{66.7\%}             & 67.4\%             \\
\hline
\end{tabular}
\label{tab_2}
\end{center}}
\end{table}

\begin{figure}[htbp]
  \centering
  \begin{minipage}[t]{0.22\linewidth}
    \centering
    \centerline{\includegraphics[width=0.9\linewidth]{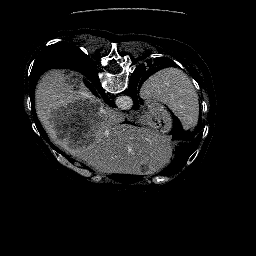}}
    \caption{original}
    \label{fig_15}
  \end{minipage}
  \begin{minipage}[t]{0.22\linewidth}
    \centering
    \centerline{\includegraphics[width=0.9\linewidth]{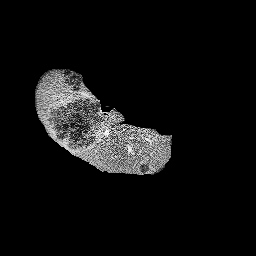}}
    \caption{liver}
    \label{fig_16}
  \end{minipage}
  \begin{minipage}[t]{0.22\linewidth}
    \centering
    \centerline{\includegraphics[width=0.9\linewidth]{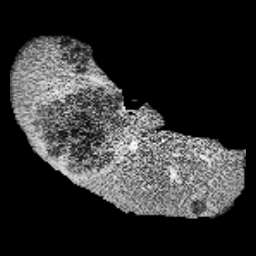}}
    \caption{resized}
    \label{fig_17}
  \end{minipage}
  \begin{minipage}[t]{0.22\linewidth}
    \centering
    \centerline{\includegraphics[width=0.9\linewidth]{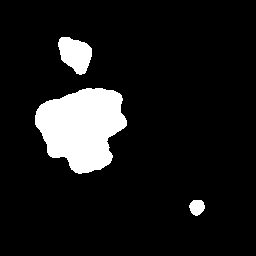}}
    \caption{tumor}
    \label{fig_18}
  \end{minipage}
\end{figure}

Finally, ``Fig.~\ref{fig_15}'' -- ``Fig.~\ref{fig_18}'' is an example of our segmentation on LiTS. ``Fig.~\ref{fig_15}'' is the original image in LiTS. ``Fig.~\ref{fig_16}'' is the liver region predicted by our liver segmentation ES-UNet++. ``Fig.~\ref{fig_17}'' resizes the liver region. ``Fig.~\ref{fig_18}'' is the tumor region of the liver area that is predicted by our tumor segmentation ES-UNet++.

\section{Conclusion}\label{5}

In this paper, we have proposed a variation of UNet++, named ES-UNet++, intended to improve the accuracy in liver and tumor segmentation. First, we discuss the relationship between encoding feature supervision and deep supervision by constructing ES-U-Net and comparing it with U-Net$^e$, which is a previous version of UNet++ without deep supervision. It is found that two models have similar dice scores for liver segmentation and both of them serve to utilize information in shallow layers. Subsequently, a cascaded ES-UNet++ is evaluated for liver and tumor segmentation respectively using a combined loss function of cross-entropy and dice. Through experiments with six versions of UNet++, which are UNet++, bottleneck feature supervision UNet++, encoding feature supervision UNet++, normal supervision UNet++, and deep supervision UNet++, we discovered that the ES-UNet++ achieves the highest dice score for both liver and tumor segmentation, has a huge improvement to model pruning operations and also accelerate convergence.

\bibliography{references}

\end{document}